# Centralized "big science" communities more likely generate non-replicable results

Valentin Danchev[1,2,3]*, Andrey Rzhetsky[3,4,5], and James A. Evans[1,2,3]*

[1] *Department of Sociology, University of Chicago, 1126 E 59th St, Chicago, IL 60637, USA*

[2] *Knowledge Lab, University of Chicago, 5735 South Ellis Avenue, Chicago, IL 60637, USA*

[3] *Computation Institute, University of Chicago, 5735 South Ellis Avenue, Chicago, IL 60637, USA*

[4] *Departments of Medicine and Human Genetics, University of Chicago, 5841 S. Maryland Ave., Chicago, IL 60637, USA*

[5] *Institute for Genomic and Systems Biology, 900 East 57th Street, Chicago, IL 60637, USA*

*Please address correspondence to vdanchev@uchicago.edu and jevans@uchicago.edu.

**Abstract**

Growing concern that most published results, including those widely agreed upon, may be false are rarely examined against rapidly expanding research production. Replications have only occurred on small scales due to prohibitive expense and limited professional incentive. We introduce a novel, high-throughput replication strategy aligning 51,292 published claims about drug-gene interactions with high-throughput experiments performed through the NIH LINCS L1000 program. We show (1) that unique claims replicate 19% more frequently than at random, while those widely agreed upon replicate 45% more frequently, manifesting collective correction mechanisms in science; but (2) centralized scientific communities perpetuate claims that are less likely to replicate even if widely agreed upon, demonstrating how centralized, overlapping collaborations weaken collective understanding. Decentralized research communities involve more independent teams and use more diverse methodologies, generating the most robust, replicable results. Our findings highlight the importance of science policies that foster decentralized collaboration to promote robust scientific advance.

**Introduction**

Concern over reliability (*1*) and reproducibility (*2, 3*) in science calls into question the cumulative process of building on prior results published by others. In a publication environment that rewards novel findings over verifications (*4, 5*), scientists remain uncertain about the published claims they assemble into new research designs and discoveries. In this paper, we demonstrate a claim replication strategy that repurposes high-throughput experiments to evaluate the replication likelihood for tens of thousands of scientific claims curated from a wide range of research articles.

Our strategy builds on the synergy of two advances. First, databases of empirical claims on many topics, ranging from material science to biomedicine (*6, 7*), are extracted from literature and linked to digital archives of scientific articles such as MEDLINE and Web of Science, making possible systematic analysis of features that characterize the provenance of a scientific claim, such as where, when, who, and how many experiments provided findings for and against it (*8*). Second, high-throughput experiments driven by consistent, programmable robots have become increasingly widespread, initially in genome sequencing (*9*) but extending to many domains, including drug-gene interactions. These experiments provide large-scale biomedical data enabling simultaneous estimation of the replication likelihood for many prior published claims.

Replication failures are typically attributed to systemic bias in a publication system that favors positive results (*1*). This incentivizes questionable research choices such as *p*-hacking (*10, 11*), "flexible" data analysis (*12*), low statistical power (*13*), selective reporting ("the file drawer problem") (*14*), and confirmation bias (*15*), which generate false results unlikely to replicate in future experiments (*12*).

Here we investigate the social and methodological structure of scientific and methodological communities that coalesce around scientific claims. What network dependencies linking scientists and their experimental methods produce the most robust, replicable claims? We hypothesize that an empirical claim tested by a *decentralized community of non-overlapping teams* with diverse training and prior knowledge, using distinct methods under varying experimental conditions will produce robust knowledge, likely to replicate in future experiments. By contrast, increased collaboration (*16*), growing teams (*17*), star scientists (*18, 19*), and expensive, shared equipment trace a pathway to "big science" (*20*) approaches to knowledge production (Fig. S1), which at the extreme involve integrated collaboration of an entire community around massive, singular research efforts like the Human Genome Project and the Laser Interferometer Gravitational-Wave Observatory (LIGO) (*20-22*). We hypothesize that a claim tested by a *centralized or densely connected community* involving repeated collaborations and a narrow range of methods, knowledge, and conditions is fragile and less likely to replicate than those produced by many decentralized or independent labs, while acknowledging that some claims can only be explored through massive, integrated collaboration.

Prior studies simulated the effect of network structure on scientific outcomes, suggesting that independent labs should more likely arrive at truth than a connected network of scientists, which can more easily propagate early, false results (*23-25*). A more recent experiment demonstrates that decentralized network ties, rather than independence, most likely improves the wisdom of crowds (*26*). Scientific findings could also be more or less robust as a function of diverse methods (*27, 28*) used to corroborate them or distinct theories used to motivate them. Consider Jean Perrin's use of multiple experimental techniques and theories to precisely determine Avogadro's number (*29*).

We represent the provenance of a scientific claim with a multilayer network (*30, 31*), where each network node depicts a research article reporting a finding about that claim. Articles are linked based on whether they (1) agree and the degree to which they share (2) scientific collaborators, (3) empirical methodologies, and (4) references to prior work (see Fig. 2A). This network representation allows us to analyse the replication likelihood for each claim as a function of the epistemic, social, technical and intellectual dependencies that underlie it.

**High-throughput Claim Replication Strategy**

We examined a corpus of 51,292 scientific claims about chemical/drug-gene/mRNA interactions in human systems. We compiled the corpus by using claims about directed drug-gene interactions curated from biomedical publications in the most comprehensive database, the Comparative Toxicogenomics Database (CTD) (*6*), recording over one million published claims regarding drug-gene interactions. Each scientific claim is a triple of drug, gene, and interaction effect. For comparability with high-throughput experiments, we selected interaction effects in which a drug "increases expression" or "decreases expression" of an mRNA in humans (effect magnitudes were not recorded). The database identifies source articles in which the finding is claimed (i.e., PubMed ID), which enabled examination of article content (e.g., methods) and metadata features (e.g., authors).

To estimate replication likelihood, we map our corpus of drug-gene claims to high-throughput experimental data from the NIH LINCS L1000 program. The program generated 1.3M gene expression profiles from 42,080 chemical and genetic perturbagens across cell lines, time points, and doses (*32*). We used profiles induced by chemical perturbagens, amounting to 19,811 small molecule compounds (including FDA approved drugs). To measure relative gene expression resulting from chemical perturbations, the LINCS L1000 team computed robust z-scores that represent differential expression signatures. We combined the z-scores across signatures using a bootstrapped modification of Stouffer's method (*33, 34*) to generate drug-gene interaction triples of (i) drug, (ii) gene (mRNA), and (iii) combined $z$-score indicating experimental effect size and effect direction. We matched drug-gene interaction triples from the LINCS L1000 experiments to triples in CTD, and found 51,292 drug-gene interaction claims at the intersection, corresponding to 60,159 supportive and 4,253 opposing findings from the literature, annotated from 3,363 scientific papers.

We distinguish between robustness to social, methodological and intellectual dependencies and robustness to experimental conditions (*27*) like time, dosages, cells, or tissues, which delimit the potential for a drug-gene interaction to generalize. In all subsequent models, we control for the variability of drug-gene interactions in LINCS L1000 to disentangle features of experimental and biological heterogeneity from global patterns of (in)dependence.

Our high-throughput replication strategy evaluates reliability of the overall scientific claim rather than of any particular finding. The latter would require precise reproduction of the original research following the original experimental protocol (*35*). Nevertheless, collective agreement across many articles may serve as evidence of a generalized claim robust to any one technique (*28, 36*), scientist, research group, or biological setting (e.g., cell line), thereby justifying increased confidence in the claim's widespread scientific and biomedical relevance.

**Results**

We observe a long-tailed distribution of published findings in support of a given scientific claim (Fig. 1A). Most claims are supported by findings in one (89%) or two (8%) articles, while few appear in many articles. The distribution of experimental effect sizes over drug-gene interactions in LINCS L1000 decays much more slowly (Fig. 1D).

Published findings widely agree on the direction of drug-gene interactions (93%) (Fig. 1A). In contrast, the majority of interaction effect-sizes (59%) in LINCS L1000 across experimental conditions are statistically undetermined (Fig. 1D). We find that consensus in the literature is positively associated with social dependencies. A pair of findings about a drug-gene interaction reported in papers with shared author(s) are significantly more likely to agree in the direction of the effect (0.989; 95% CI: 0.984, 0.993; 2,514 pairs of papers) than findings reported in socially independent papers (0.889; 95% CI: 0.884, 0.893; 20,846 pairs of papers) (Fig. S2).

We also find that generalizable interactions, which converge across experimental conditions in LINCS L1000, are more likely to replicate published claims ($RR = 0.60$) than interactions that vary across conditions ($RR = 0.52$), which are indistinguishable from random ($RR = 0.501$) in our corpus (Fig. 1E). We define the replication success ($R = 1$; 0 otherwise) of a claim as the effect size (combined z-score) in LINCS L1000 matching the direction (increase or decrease) of the effect claimed in literature (see SM).

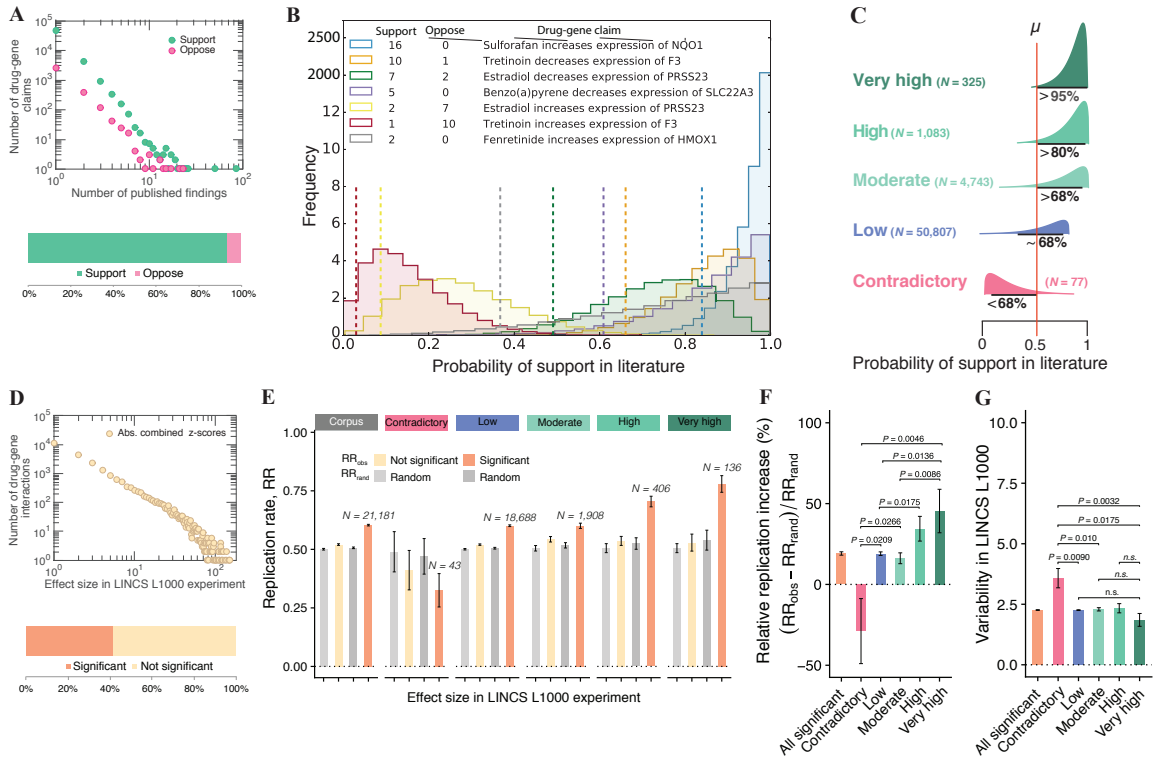

**Fig. 1. Aligning drug-gene claims from the literature with interactions from high-throughput experiments to evaluate replication likelihood.** (**A**) On top, the distribution of drug-gene claims against the number of supporting and opposing findings in the literature on a log-log scale. On bottom, the proportion of supporting and opposing findings on aggregate. (**B**) Posterior distributions of support in the biomedical literature for a sample of drug-gene claims. For each claim, we summarize the probability of support (dashed line) using the lower bound on the 95% posterior credible intervals (PCI). (**C**) Typology of drug-gene claims from the literature on the basis of claim's distribution of support compared to the null value of 0.5. Number of claims included in a category provided in brackets. (**D**) Effect sizes (combined $z$-scores) for drug-gene interactions derived from LINCS L1000 experiments. On top, the distribution of drug-gene interactions against absolute experimental effect size in LINCS L1000 (log-log scale). On bottom, the proportion of significant and insignificant interactions in LINCS L1000. We established significance at the 0.05 level by bootstrapping (10,000 iterations) the combined z-score for each of the 51,292 interactions sampling from experiments across different cell lines, durations, and dosages. (**E**) Observed, $RR_{obs}$, versus expected, $RR_{rand}$, replication rate (see SM) for drug-gene claims differentiated by significance in LINCS L1000 and support from literature. (**F**) Relative replication increase for each type of claim, indicating the percentage increase of $RR_{obs}$ relative to random $RR_{rand}$ or $100 \times \frac{RR_{obs} - RR_{rand}}{RR_{rand}}$. (**G**), Variability (coefficient of variation) in LINCS L1000 across cell lines, durations, and dosages for each type of claim. We established statistical significance using a nonparametric bootstrap test with 100,000 iterations. Error bars represent SEM.

A central concern is whether the replication problem applies only to novel and rare claims or if it also afflicts widely supported results, as recently hypothesised (*37, 38*). To integrate collective support in the literature over multiple claim sources, we design a binomial Bayesian model with a uniform prior that accommodates skewed distributions (*39*) like that of findings per claim we observed. To estimate the probability of support for each drug-gene claim, we assume that the number of supportive published findings $\gamma$ in $n$ findings about that claim is drawn from a binomial distribution, $p(\gamma|\theta) \sim \text{Bin}(\gamma|n, \theta)$ (see SM). The model allocates higher probability to scientific claims unanimously supported in a large number of articles and lower probability to infrequent and disputed claims (Fig. 1B).

We find a strong relationship between scientific support in the literature $L_{\text{SUPT}}$ and the probability of replication success, ranging from 0.507 (95% CI: 0.492, 0.522) for claims with the least support to 0.705 (95% CI: 0.665, 0.743) for claims with the largest, controlling for biological and experimental variability in LINCS L1000 (Fig. S3A). When we consider only significant drug-gene interactions (Fig. S3B), which generalize across experimental conditions, the probability of replication for top-supported claims reaches 0.798 (95% CI: 0.751, 0.845).

To examine the hypothesis that convergent drug-gene claims more likely replicate than those claimed rarely or disputed in literature, we categorized claims on the basis of their distribution of support (Figs. 1C and S4; see SM). We found that claims with "Very high" (*RRI* = 45.4%; SEM ± 13.5%) and "High" (*RRI* = 34.5%; SEM ± 7.6%) support in the biomedical literature are significantly more likely to replicate in high-throughput experiments than claims with "Low" (*RRI* = 19%; SEM ± 1.1%) and "Moderate" (RRI = 16.2%; SEM = 3.3%) support (Fig. 1F). The replication of "Low" and "Moderate" support claims is consistent with estimates (~11–25%) from recent biomedical reproducibility studies (*2, 3*), but those receiving support from multiple sources is much higher. "Contradictory" claims are significantly less likely to replicate than random. They are also associated with greater experimental variability (Fig. 1G), indicating that collective disagreements among findings truthfully signal unstable drug-gene interactions with low replication likelihood. These results suggest that findings reported in a single scientific article may be fallible, but viewed as a system of converging and diverging findings, science exhibits collective correction.

Independent and decentralized sources of evidence should increase claim robustness. Figure 2A presents a schematic of a multilayer network $M = (V_M, E_M, L)$ for a claim. In each network layer $L$, nodes $V$ are scientific papers and edges $E$ between pairs of papers represent either a binary relationship of agreement ($L_1$) or the amount of overlap measured with the Jaccard coefficient, between the two papers' sets of authors ($L_2$), methodologies ($L_3$), and references to prior literature ($L_4$). Our independence scores for social, methodological, and prior knowledge range between 0 (dependent) and 1 (independent), and can be viewed as the probability that any two randomly chosen findings about a claim are obtained by disconnected sets of authors, methods, and references, respectively (see SM). Figure 2B-C present schematics of bipartite networks of authors connected to the papers they publish. To quantify the centralization of research communities $C$, we compute the Gini coefficient of the authors' degree distribution (see Fig. 2D-E).

We examine the impact of network dependencies on claim replicability by fitting logistic regression models that predict replication success against the measures described above and defined for each claim, using a sub-corpus of 2,493 claims with determined direction of the drug-gene effect in the literature ($L_{\text{SUPT}} = 68\% \text{ PCI} > \mu < 68\% \text{ PCI}$) and in LINCS L1000. Figure 2F shows that the odds ratios (OR) of replication increase substantially with the increase of scientific support in the literature $L_{\text{SUPT}}$ (OR 23.20; 95% CI: 9.08, 59.3), social independence $S_{\text{IND}}$ (OR 6.31; 95% CI: 4.07, 9.79), methodological independence $M_{\text{IND}}$ (OR 6.30; 95% CI: 3.44, 11.53), and prior knowledge independence $K_{\text{IND}}$ (OR 5.53; 95% CI: 2.58, 11.84). By contrast, claim replication decreases with centralization $C$ (OR 0.36; 95% CI: 0.27, 0.48). Our estimates indicate that claim robustness, defined here as multiple decentralized evidence, increases replication success (Fig. 2G-M). When all predictors are modelled simultaneously (Fig. 2F), however, centralization and support in the literature account for the rest, suggesting that centralized academic training and extensive collaboration mediates the application of research techniques and attention to prior research.

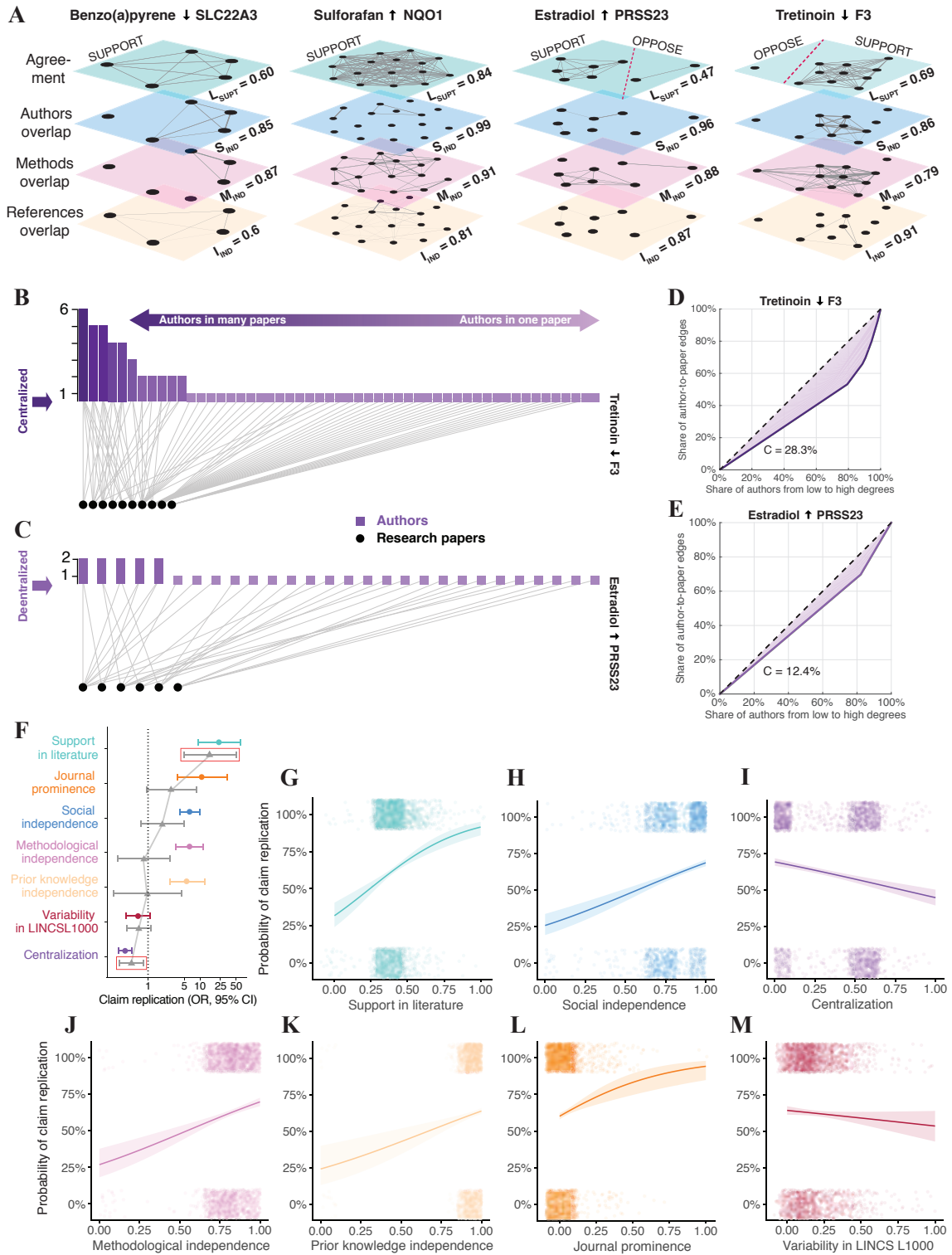

**Fig. 2. Multilayer networks of social, methodological, and prior knowledge dependences predict replication likelihood.** (A) Examples of multilayer claim networks. Each layer $L$ represents a co-paper network: nodes are scientific papers; pairs of papers are connected by an unweighted edge if they agree on the effect direction ($L_1$),

and a weighted edge proportional to the overlap (Jaccard coefficient or JC) between the two papers' authors ($L_2$), methodologies ($L_3$), and references to prior literature ($L_4$). Dashed red line in the agreement layer indicates opposing edges. For clarity, we only plot edges above the mean JC value for overlapping methods. (**B-C**) Example of centralized and decentralized bipartite networks. Edges connect authors (rectangles) to papers they published (circles). Centralization $C$ is the Gini coefficient of the authors' degree distribution. (**D-E**) Lorenz curves showing different level of centralization, corresponding to the examples in B and C. (**F**) Logistic regression models with replication success as the response variable and predictors modelled (top) independently and (bottom) simultaneously. Predictors are rescaled $\frac{x_i - \min(x)}{\max(x) - \min(x)}$ for comparability. (**G-M**) Predicted probabilities (PP) of claim replication $PP(R=1) = \frac{\exp(\beta_0 + \beta_1 X)}{1 + \exp(\beta_0 + \beta_1 X)}$ for logistic models of independent predictors with 95% error bounds.

An alternative explanation of replication success is the biological tendency for some drug-gene interactions to generalize across conditions and so replicate in future experiments. Figure 2F shows that experimental variability has a small, insignificant negative effect, and that support from multiple, decentralized teams are much more informative predictors of replication success.

Our combined model also controls for journal prominence *J*, which we measure with journal eigenfactor (*40*), a score that credits journals receiving many citations from highly cited journals. Claim replication increases with journal prominence (Figs. 2F and 2L), but prominent journals are responsible for only a tiny fraction of all claims, which warrants our evaluation strategy and the practice of extracting and archiving findings from a wide range of journals (*6*).

Figure 3 shows that by accounting for scientific support, centralization, and social independence, we can identify claims with high replication probability. Claims supported by many publications have about 45 percent higher probability to replicate when investigated by decentralized versus centralized communities (Figs. 3A and S8). Even if a claim garners wide support, if studied exclusively by a centralized scientific community, the claim is indistinguishable in replication probability from a claim reported in a single paper. This suggests a condition under which collective correction in science is undermined, when one or several scientists exercise disproportionate influence on team assembly and research across multiple investigations of a claim. Claims robust to multiple, socially independent investigations have a 55 percent higher probability to replicate than those studied by small clusters of overlapping collaborators (Figs. 3B and S9). These effects all account for the experimental and biological variability of drug-gene interactions.

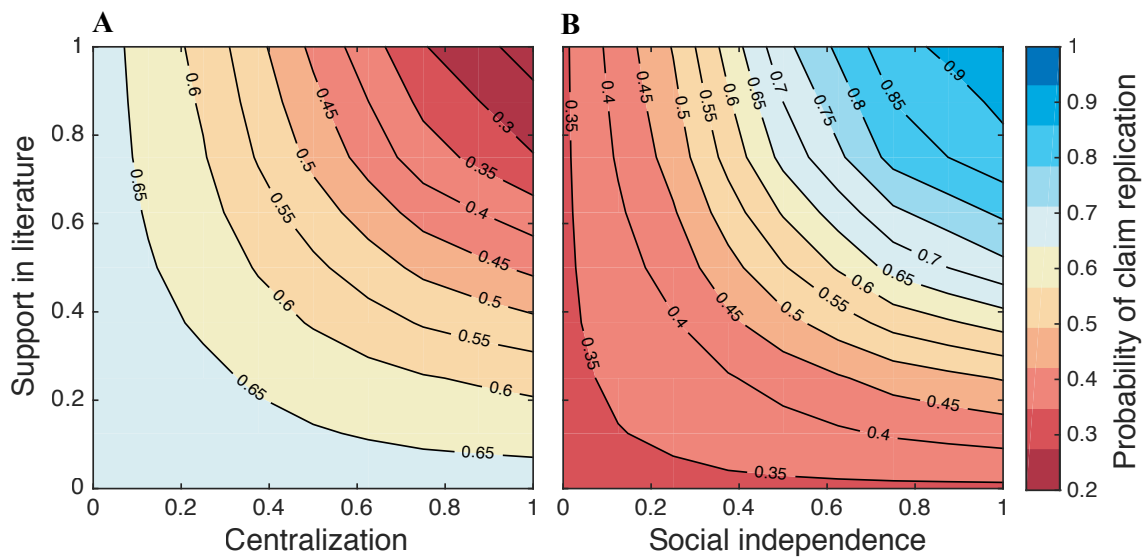

**Fig. 3. Contour plots of the probability of claim replication based on network dependencies and support in the literature.** (**A**) The probability of claim replication is greater in decentralized communities of scientists, after controlling for variability in LINCS L1000 across cell lines, durations, and dosages. (**B**) The probability of claim replication increases with support in the literature and social independence, after controlling again for experimental variability.

**Discussion**

Extensive (*22*) and often overlapping scientific collaboration (*16*), along with cumulative advantage processes that create central, star scientists (*19*), produce centralized "big science" communities with dense methodological and intellectual dependencies. Such communities tend to publish fragile findings. Our research points to the importance of science policies that foster competition and decentralized collaboration to promote robust and replicable scientific advance. Our analysis demonstrates the utility of large-scale experiments coupled with enriched article content and metadata to diagnose the replicability of published research. Our findings suggest a calculus for evaluating the system-level trade-off between investments in robust, replicable knowledge, which comes at the price of larger, but weaker, preliminary insight.

**References**

1. J. P. Ioannidis, Why most published research findings are false. *PLoS Med* **2**, e124 (2005).
2. F. Prinz, T. Schlange, K. Asadullah, Believe it or not: how much can we rely on published data on potential drug targets? *Nat Rev Drug Discov* **10**, 712–712 (2011).
3. C. G. Begley, L. M. Ellis, Drug development: Raise standards for preclinical cancer research. *Nature* **483**, 531–533 (2012).
4. B. A. Nosek, J. R. Spies, M. Motyl, Scientific Utopia: II. Restructuring incentives and practices to promote truth over publishability. *Perspectives on Psychological Science* **7**, 615–631 (2012).
5. B. Alberts *et al.*, Self-correction in science at work. *Science* **348**, 1420 (2015).
6. A. P. Davis *et al.*, The Comparative Toxicogenomics Database: update 2017. *Nucleic Acids Research* **45**, 972–978 (2017).
7. A. H. Wagner *et al.*, DGIdb 2.0: mining clinically relevant drug–gene interactions. *Nucleic acids research*, gkv1165 (2015).
8. J. A. Evans, J. G. Foster, Metaknowledge. *Science* **331**, 721 (2011).
9. International Human Genome Sequencing. Consortium, Initial sequencing and analysis of the human genome. *Nature* **409**, 860 (2001).
10. M. L. Head, L. Holman, R. Lanfear, A. T. Kahn, M. D. Jennions, The extent and consequences of p-hacking in science. *PLoS Biol* **13**, e1002106 (2015).
11. U. Simonsohn, L. D. Nelson, J. P. Simmons, P-curve: a key to the file-drawer. *Journal of Experimental Psychology: General* **143**, 534 (2014).
12. J. P. Simmons, L. D. Nelson, U. Simonsohn, False-Positive Psychology. *Psychological Science* **22**, 1359–1366 (2011).
13. E. Dumas-Mallet, K. S. Button, T. Boraud, F. Gonon, M. R. Munafò, Low statistical power in biomedical science: a review of three human research domains. *Royal Society Open Science* **4**, 160254 (2017).
14. R. Rosenthal, The file drawer problem and tolerance for null results. *Psychological Bulletin* **86**, 638–641 (1979).
15. R. Nuzzo, Fooling ourselves. *Nature* **526**, 182–185 (2015).
16. R. Guimerà, B. Uzzi, J. Spiro, L. A. N. Amaral, Team Assembly Mechanisms Determine Collaboration Network Structure and Team Performance. *Science* **308**, 697 (2005).
17. S. Wuchty, B. F. Jones, B. Uzzi, The Increasing Dominance of Teams in Production of Knowledge. *Science* **316**, 1036 (2007).
18. R. K. Merton, The Matthew effect in science. *Science* **159**, 56–63 (1968).
19. P. Azoulay, T. Stuart, Y. Wang, Matthew: Effect or Fable? *Management Science* **60**, 92-109 (2013).
20. D. J. de Solla Price, *Little science, big science*. (Columbia University Press, New York, 1963).
21. A. M. Weinberg, Impact of Large-Scale Science on the United States. *Science* **134**, 161 (1961).
22. D. Hicks, J. S. Katz, Science policy for a highly collaborative science system. *Science and Public Policy* **23**, 39–44 (1996).

23. N. Payette, in *Models of Science Dynamics: Encounters Between Complexity Theory and Information Sciences,* A. Scharnhorst, K. Börner, P. van den Besselaar, Eds. (Springer, Berlin, Heidelberg, 2012), pp. 127–157.
24. Kevin J. S. Zollman, The Communication Structure of Epistemic Communities. *Philosophy of Science* **74**, 574–587 (2007).
25. J. Lorenz, H. Rauhut, F. Schweitzer, D. Helbing, How social influence can undermine the wisdom of crowd effect. *Proceedings of the National Academy of Sciences* **108**, 9020–9025 (2011).
26. J. Becker, D. Brackbill, D. Centola, Network dynamics of social influence in the wisdom of crowds. *Proceedings of the National Academy of Sciences* **114**, E5070-E5076 (2017).
27. W. G. Kaelin Jr, Common pitfalls in preclinical cancer target validation. *Nat Rev Cancer* **17**, 425–440 (2017).
28. W. C. Wimsatt, in *Characterizing the Robustness of Science: After the Practice Turn in Philosophy of Science,* L. Soler, E. Trizio, T. Nickles, W. Wimsatt, Eds. (Springer Netherlands, Dordrecht, 2012), pp. 61–87.
29. W. Salmon, *Scientific explanation and the causal structure of the world*. (Princeton University Press, Princeton, New Jersey, 1984).
30. M. Kivelä *et al.*, Multilayer networks. *Journal of Complex Networks* **2**, 203–271 (2014).
31. J. F. Padgett, C. K. Ansell, Robust Action and the Rise of the Medici, 1400–1434. *American Journal of Sociology* **98**, 1259–1319 (1993).
32. A. Subramanian *et al.*, A Next Generation Connectivity Map: L1000 Platform and the First 1,000,000 Profiles. *Cell* **171**, 1437–1452.e1417.
33. M. C. Whitlock, Combining probability from independent tests: the weighted Z-method is superior to Fisher's approach. *Journal of evolutionary biology* **18**, 1368–1373 (2005).
34. D. S. Himmelstein *et al.*, Systematic integration of biomedical knowledge prioritizes drugs for repurposing. *eLife* **6**, e26726 (2017).
35. T. M. Errington *et al.*, An open investigation of the reproducibility of cancer biology research. *Elife* **3**, e04333 (2014).
36. B. A. Nosek, T. M. Errington, Making sense of replications. *eLife* **6**, e23383 (2017).
37. S. B. Nissen, T. Magidson, K. Gross, C. T. Bergstrom, Publication bias and the canonization of false facts. *eLife* **5**, e21451 (2016).
38. R. McElreath, P. E. Smaldino, Replication, Communication, and the Population Dynamics of Scientific Discovery. *PLOS ONE* **10**, e0136088 (2015).
39. C. Davidson-Pilon, *Bayesian Methods for Hackers: Probabilistic Programming and Bayesian Inference*. (Pearson Education, 2015).
40. C. T. Bergstrom, J. D. West, M. A. Wiseman, The Eigenfactor™ Metrics. *The Journal of Neuroscience* **28**, 11433 (2008).
41. J. Kruschke, *Doing Bayesian Data Analysis: A Tutorial with R, JAGS, and Stan*. (Elsevier Science, 2014).
42. A. Gelman *et al.*, *Bayesian Data Analysis*. (Taylor & Francis, ed. Third edition, 2014).

43. M. Kivelä, Multilayer Networks Library for Python (Pymnet), available at http://bitbucket.org/bolozna/multilayer-networks-library. 2017.
44. S. Wasserman, K. Faust, *Social network analysis: methods and applications*. (Cambridge University Press, Cambridge, 1994), pp. 825.
45. A. Bandrowski *et al.*, The Ontology for Biomedical Investigations. *PLOS ONE* **11**, e0154556 (2016).
46. A. Gonzalez-Beltran, P. Rocca-Serra, O. Burke, S.-A. Sansone. (Available online: http://bioportal.bioontology.org/ontologies/STATO, Last accessed on 07 November 2017, 2016).
47. M. M. Kessler, Bibliographic coupling between scientific papers. *American Documentation* **14**, 10–25 (1963).
48. B. Peirson, R. Erick, e. al. (http://diging.github.io/tethne/, 2017).
49. A. Hagberg, P. Swart, D. S Chult, "Exploring network structure, dynamics, and function using NetworkX,"  (Los Alamos National Laboratory (LANL), 2008).
50. J. Kunegis, J. Preusse, in *Proceedings of the 3rd Annual ACM Web Science Conference*. (ACM, 2012), pp. 175–184.
51. L. C. Freeman, Centrality in social networks conceptual clarification. *Soc Networks* **1**, 215–239 (1978).
52. J. M. Badham, Commentary: Measuring the shape of degree distributions. *Network Science* **1**, 213–225 (2013).
53. J. Fox, Effect displays in R for generalised linear models. *Journal of statistical software* **8**, 1–27 (2003).
54. D. Lüdecke, sjPlot: data visualization for statistics in social science. *R package version* **2.3.3** (2017).

**Acknowledgments**

We thank A. Belikov, D. Centola, E. Duede, J. P.A. Ioannidis, M. Lewis, and T. Stoeger for helpful discussions and comments, E. Demir and R. Melamed for advice on curated drug-gene databases and high-throughput experimental data, respectively, W. Catino for computational help, M. Kivelä for help with his Multilayer Networks Library for Python (Pymnet), and T. Natoli from Broad Institute for advice on NIH LINCS L1000 data. We used MEDLINE/PubMed and Web of Science data to build the co-paper networks. We thank Clarivate Analytics for supplying the Web of Science data and T. Ando for computing the journal eigenfactor scores. This work was supported by DARPA's Big Mechanism research program grant 14145043, NSF SciSIP grant 1158803, and AFOSR grant FA9550-15-1-0162.

# Supplementary Materials

**List of supplementary content:**

    Materials and Methods
    Data Analysis
    Figures
    Tables

**Materials and Methods**

DATA

High-throughput drug-gene interactions
We used the Library of Integrated Network-based Cellular Signatures (LINCS) Phase I L1000 data set to estimate claim replication (*32*). The data set measures the expression of 978 landmark genes treated with wide range of perturbagens across cell lines, time points, and doses (concentration), resulting in approximately 1.3M profiles. The experimental profiles are aggregated to 473,647 signatures, as represented in Level 5 data we used to perform our analysis. The landmark genes are subsequently used to infer gene expressions for the remaining genes in the human genome. In addition to the 978 landmark genes, we consider 9,196 for which the LINCS L1000 project estimated to be well inferred, resulting in 10,174 Best INferred Genes (BING) in total. With respect to perturbagen types, we used the set of small-molecule compounds (19,811 compounds), which includes a subset of approximately 1,300 FDA-approved drugs. We accessed the data file GSE70138_Broad_LINCS_Level5_COMPZ_n118050x12328_2017-03-06.gctx.gz and metadata from the GEO depository at https://www.ncbi.nlm.nih.gov/geo/query/acc.cgi?acc=GSE70138.

Published drug-gene interactions
We analysed curated data about published interactions between chemicals/drugs and genes/mRNAs released by the Comparative Toxicogenomics Database (CTD) (*6*) on June 7 2016. (The current CTD data release is available here: http://ctdbase.org/downloads/.) To align triples of drug, gene, and interaction effect in CTD to corresponding triples in the experimental signatures from LINCS L1000, we performed the following procedures. First, we selected drug-gene interactions about Homo sapiens, comprising approximately 40% of the CTD data. Second, CTD reports the form of the gene (e.g., mRNA, protein) that is implicated, and we selected only mRNA as LINCS L1000 measures gene expression at the mRNA level. Third, we mapped chemical names and Entrez gene IDs in the CTD to perturbagen names and Entrez gene IDs in LINCS L1000. Fourth, to ensure comparability to the LINCS L1000 signatures, we selected drug-gene interactions with a single interaction effect, either "decreases expression" or "increases expression", defining the direction of the effect that chemical/drug manifests on a gene/mRNA. Note that we do not consider complex interactions with multiple, nested effects in our analysis.

Likewise, interactions for which the direction of the effect is not specified, such as "affects binding", are not considered. The resulting corpus at the intersection of LINCS L1000 and CTD comprises 51,292 drug-gene claim combinations of 605 unique drugs and 9,123 unique genes, annotated from 3,363 scientific articles.

METHODS

Combining LINCS L1000 effect sizes
The LINCS L1000 team applied a moderated Z-scoring procedure to estimate weighted average signatures. For each drug-gene interaction, we computed a combined gene signature using Stouffer's method (*33, 34*) $Z = \frac{\sum_{i=1}^{k} Z_i}{\sqrt{k}}$, where $Z_i$ is an experimental z-score and $k$ is the number of signatures. Using this procedure, we aggregated experiments about drug-gene interactions across cell lines, dosages, and durations into a combined effect size (z-score). Some drug-gene interactions generalize across conditions, whereas others are context specific. We define as generalized drug-gene effect sizes that are significant at the 0.05 level. To measure significance, for each drug-gene interaction, we performed 10,000 bootstrap iterations on the combined z-score, sampling from experiments performed under different cell lines, dosages, and durations. Conventionally, if the 95% confidence intervals do not contain the null value of 0, we consider a drug-gene interaction as generalized.

Measuring variability in LINCS L1000
To measure how each drug-gene interaction varies across cell lines, dosages, and durations in LINCS L1000, we computed the coefficient of variation. The coefficient of variation (CI) is the ratio of the standard deviation to the absolute value of the mean: $CI = \frac{\sigma}{\text{abs}(\mu)}$. In comparison to scale-dependent measures of variability, CI is a normalized measure that allows us to make comparisons across drug-gene interactions.

Modelling claims' support in literature
We apply a binomial Bayesian model (*39, 41, 42*) to estimate the probability of scientific support $\theta$ for each drug-gene claim given the number of supportive findings $\gamma$ in all findings $N$ reported in CTD about a particular drug-gene interaction. We assume that the prior distribution of $\theta$ is uniform on the interval [0,1]: $\theta_i \sim \text{Uniform}(\min = 0, \max = 1)$. This uninformative prior considers all possible probability support values $\theta$ as equally possible and is suitable for the skewed distribution of findings over drug-gene claims. We assume that the number of findings in support of a drug-gene claim $\gamma$ follows a binomial distribution $p(\gamma|N, \theta) \sim \text{Binomial}(\gamma|N, \theta)$ where $\text{Binomial}(\gamma|N, \theta) \sim \binom{N}{\gamma} \theta^\gamma (1 - \theta)^{N-\gamma}$. Then, the posterior density for $\theta$ is $p(\theta|\gamma) \propto \theta^\gamma (1 - \theta)^{N-\gamma}$. We approximated the posterior density of $\theta$ for each drug-gene claim by performing 10,000 Markov chain Monte Carlo (MCMC) sampling iterations (2,500 burn in iterations) for each drug-gene claim using the Metropolis–Hastings MCMC sampler implemented in the PyMC package (version 2.3.6) for Python. To improve convergence, we approximate the maximum posterior (MAP) before running the MCMC sampler (*39*).

Typology of drug-gene claims
To avoid setting arbitrary thresholds to categorize claims by their support in literature, we used the posterior distributions from our Bayesian model of support. For each claim, we estimated the overlap between the posterior credible intervals (PCI) and the null value of $\mu = 0.5$ (Fig. 1C): *Very high support* claims ($L_{SUPT}$ 95% PCI exceeds $\mu$) yield agreement from multiple papers (~7 papers on average), amounting to 325 claims that are supported by 2,241 findings, but only opposed by 21; *High support* claims (80% PCI $\geq L_{SUPT}$ < 95% PCI exceeds $\mu$) yield agreement from 3 papers on average ($N = $ 1,083 claims, supported and opposed in 3,525 and 42 articles, respectively); *Moderate support* claims (68% PCI $\geq L_{SUPT}$ < 80% PCI exceeds $\mu$) yield agreement from 2 papers on average ($N = 3,743$ claims, supported and opposed in 7,557 and 38 articles, respectively); *Low support* claims (68% PCI $\geq L_{SUPT}$ < 80% PCI contains $\mu$) are overwhelmingly supported by a single paper or opposed by virtually the same number of papers that support them such that the direction of the effect is undetermined ($N = 46,064$ claims, supported and opposed in 46,735 and 3,668 articles, respectively). *Contradictory claims* ($L_{SUPT}$ 68% PCI is smaller than $\mu$) generate lower support than expected as a greater number of papers reported findings in the opposite direction ($N = 77$ claims, supported and opposed in 101 and 484 articles, respectively) (Fig. S4).

Null model for estimating relative replication increase
We define replication as a binary variable $R \in \{0,1\}$, where $R = 1$ indicates that the direction (increase or decrease) of the effect claimed in literature is matched by the combined z-score in LINCS L1000, and $R = 0$ otherwise. To estimate the amount of relative replication increase (*RRI*), we first empirically established the expected or random replication rate $RR_{rand}$ by iteratively randomizing LINCS L1000 drug-gene interactions. For each randomization, we computed the replication rate for all significant drug-gene claims ($N = 21,181$) as well as for the collection of very high support ($N = 136$), high support ($N = 406$), moderate ($N = 1,908$), low ($N = 18,688$), and contradictory ($N = 43$) claims against the drug-gene interactions from LINCS L1000 that randomly matched them. We repeated the estimation procedure 100,000 times to generate randomized distributions, mean randomized replication rates $RR_{rand}$, and standard errors of the $RR_{rand}$ (or SEM) for each claim type. This randomization model takes into account biases in the distribution of published claims, as when disproportionately more claims of a certain direction, either "increasing" or "decreasing", are present. This allows us to conservatively estimate the expected replication rate on the basis of empirical data, rather than assuming a random null value of 0.5. We bootstrapped (100,000 iterations with replacement) the observed replication vector for each type of claim to generate distributions of observed replication rates, mean observed replication rates $RR_{obs}$, and the standard error of $RR_{obs}$ (or SEM). We used the randomized and bootstrapped distributions for each type of claim to estimate the percentage relative replication increase: $RRI = 100 \times \frac{RR_{obs} - RR_{rand}}{RR_{rand}}$.

Claim multilayer networks
To investigate the robustness of drug-gene claims, we represent each claim as a multilayer network $M = (V_M, E_M, L)$ (*30*). The network consists of four layers, which we plot using the Multilayer Networks Library for Python (Pymnet) (*43*). Nodes $V_M$ in the network are scientific papers. In each layer $L$, edges $E$ between pairs of papers represent either a binary relationship of agreement and disagreement about effect direction ($L_1$), or the amount of overlap between the two articles' authors ($L_2$), methods ($L_3$), and references ($L_4$). For this analysis, we exclude claims with low support as estimated via our Bayesian model because they are primarily reported by a single paper and therefore, lack independence and robustness. To examine the impact of claim robustness on the probability of replication in high-throughput experiments, we interrogated only interactions that produce a replication signal by posting a statistically significant effect size. These are claims that generalize across experimental and biological conditions in LINCS L1000. By exploring only statements having significant *agreement within* the literature and within LINCS L1000, we can directly examine the effect of social, methodological, and epistemic dependencies on shaping *agreement between* literature and high-throughput experiment. The sub-corpus of claims both supported in the literature and significant within LINCS L1000 consists of 2,493 claims, associated with 6,272 supporting and 339 opposing findings from 1,282 papers. For each claim, we create a multilayer co-paper network and model the predictive power of each layer on the replicability of a claim in the LINCS L1000 experiments.

Quantifying overlap between articles' authors, methods, and references
To quantify the amount of overlap between papers, we used the Jaccard coefficient (JC). For any two sets of paper's attributes—i.e., authors, methods, or references—$A_i$ and $A_j$, JC is the size of intersection divided by the size of the union:
$JC(A_i, A_j) = \frac{|A_i \cap A_j|}{|A_i| + |A_j| - |A_i \cap A_j|}$. The resulting quantity represents the edge weight between a pair of articles in the respective network layers of shared authors ($L_2$), methods ($L_3$), and references ($L_4$). Each drug-gene claim constitutes an undirected, multilayer network of papers connected via such weighted edges across layers (see Fig 2A).

Independence of findings
We define an independence score *IND* as the proportion of maximum possible edges (*44*) in a network layer $E_{max} = \frac{E}{n(n-1)/2}$ not present, $IND = \frac{E_{max} - W}{E_{max}}$, where $W$ is the sum over all weighted edges in a claim's respective layer of shared authors, methods, or references. The independence scores for social, methodological, and prior knowledge approaches 1 when most papers with findings in support of a claim share no common authors, methods, and references, and 0 when all papers share all of their authors, methods, and references, respectively.

Social independence
We used the MEDLINE/PubMed database to extract the set of authors for each paper. To measure the overlap between two sets of authors, we need individual identifiers for authors. Authors' name disambiguation is a common problem in research on scientific knowledge production. We used the individual identifiers based on author last name and initials. We note that because we assessed authors separately for each claim, our conservative matching procedure is very unlikely to produce false positive author linkages, and so our author co-paper network should be considered a lower bound for author co-paper density. For the sub-corpus of 2,493 claims, sourced from 1,282 papers, we estimated a mean of 6.4 authors per paper and a mean of 23.5 authors per scientific community defined here as the total number of authors that have published papers reporting a drug-gene claim. For the set of papers supporting a claim, we applied JC to measure the overlap between any two sets of papers' authors and utilized the output to compute claim's social independence $S_{IND}$ using our independence score.

Methodological independence
We compiled a controlled vocabulary of 3,074 terms (incl. synonyms) concerning methods, techniques, and experimental apparatus used in biomedical research using ontologies of biomedical investigations (*45*) and statistics (*46*). We then used the RESTful API Web Service of Europe PMC to query the methods sections from 4.4 million full text articles and extracted, on aggregate, 13,095 terms for 488 articles (38%). In parallel, for all 1,282 articles that share a drug-gene claim, we applied fuzzy matching against our vocabulary using the difflib module in Python and extracted 12,135 terms from abstracts available in MEDLINE/PubMed. We combined the outputs from the two search procedures. Then, for the set of papers supporting a claim, we again used JC to measure the overlap between any two sets of papers' methods and then employed our independence score to measure claim's methodological independence $M_{IND}$.

Prior knowledge independence
To examine whether a pair of publications is exposed to similar or dissimilar prior information, we use the notion of bibliographic coupling (*47*), i.e., the number of citations any two papers share. To compute bibliographic coupling, we used the Web of Science citation data. Out of 1,282 papers sharing a drug-gene claim with at least one other paper, we mapped 1,234 PubMed IDs to Web of Science IDs and performed bibliographic coupling on this subset using Python modules Tethne (*48*) and NetworkX (*49*). 880 of our 1,234 papers were coupled bibliographically by at least one paper. Consistent with the procedure we applied to measure shared authors and methods, we used JC to measure the overlap between any two sets of papers' citations and then employed our independence score to measure claim's prior knowledge independence $K_{IND}$.

Centralization of scientific communities
To quantify the centralization of scientific communities in the bipartite author-article networks for each claim, we employed the Gini coefficient. The Gini coefficient is used to measure the heterogeneity of distributions in social and information networks (*50*). The coefficient ranges between 0 and 1. In the context of a bipartite author-article network, the coefficient approaches 0 when all investigators author equal numbers of articles about a claim and increases to 0.3 and above (depending on the number of articles), when one investigator authors all articles and all others author only one. The Gini coefficient can be also represented as a percentage ranging from 0 to 100, as in Figure 2D-E. While other measures of network centralization are available (e.g., Freeman's centralization (*51*)), the Gini coefficient, and the Lorenz curve on which it is based, is independent from the underlying degree distribution, making it suitable for comparisons among networks with different size and mean degree (*50, 52*).

Journal prominence

To measure journal prominence, we employed the journal eigenfactor score (*40*). The eigenfactor score acts like a recursively weighted degree index by rating journals highly that receive citations from journals that are themselves highly cited. Using the Web of Science database, we computed journal eigenfactor scores for 3,162 papers (94% of all 3,363 papers in our corpus) published in 656 journals between 1995 and 2016. For the sub-corpus of 1,282 papers with shared drug-gene claims, we recovered 1,212 papers or 95% published in 496 journals. For each claim, we computed mean journal eigenfactor scores by averaging over the eigenfactor score of all journals that published a paper reporting findings in support of the claim. The distribution of mean journal eigenfactor scores (i.e., journal prominence) per claim is highly skewed (Figs. S4 and S5), indicating that claims receive overwhelmingly support from findings published in low and medium ranked journals. This highlights the value of archiving findings across a wide range of journals, as do CTD and other scientific database projects, which makes possible our large-scale evaluation of scientific output.

Shared authors and scientific agreement

To examine whether papers with shared authors are more likely to agree on the direction ("increasing expression" or "decreasing expression") of the drug-gene interaction, we identified all pairs of papers with shared or distinct authors that reported findings about the same drug-gene pair irrespective of the directionality of the effect. We found that among the 2,514 edges between papers with shared authors, 2,486 (98.9%) agreed on the direction of the effect. By comparison, among the 20,846 pairs of papers with distinct authors, 18,527 (88.9%) agreed on the direction of the effect. To examine the differences in agreement rates among papers with common versus distinct authors, we performed a two-tailed bootstrap test (100,000 samples) (Fig. S2).

Predictors of replication success

We estimate logistic regression models to predict claim replication $R$ [Replicated = 1, Non-replicated = 0] as a function of support in the literature $L_{SUPT}$, social independence $S_{IND}$, methodological independence $M_{IND}$, prior knowledge independence $K_{IND}$, centralization $C$, journal prominence $J$, and experimental variability $V$. First, for exploratory purposes, we model each variable independently (see Table S1):

$$\begin{aligned} \text{logit } P(R = 1) &= \beta_0 + \beta_1 \times L_{SUPT} \\ &\quad \beta_0 + \beta_1 \times S_{IND} \\ &\quad \beta_0 + \beta_1 \times M_{IND} \\ &\quad \beta_0 + \beta_1 \times K_{IND} \\ &\quad \beta_0 + \beta_1 \times C \\ &\quad \beta_0 + \beta_1 \times J \\ &\quad \beta_0 + \beta_1 \times V. \end{aligned}$$

Second, we model our variables simultaneously (see Table S2):

$$\text{logit } P(R = 1) = \beta_0 + \beta_1 \times L_{SUPT} + \beta_2 \times S_{IND} + \beta_3 \times M_{IND} + \beta_4 \times K_{IND} + \beta_5 \times C + \beta_6 \times J + \beta_7 \times V.$$

Third, we estimate two interaction models to examine the effect of support in the literature on replication success as a function of social independence and centralization, respectively (see Figs. 3 and S8–9):

$$\text{logit P}(R = 1) = \beta_0 + \beta_1 \times L_{\text{SUPT}} \times S_{\text{IND}} + \beta_2 \times V$$
$$\beta_0 + \beta_1 \times L_{\text{SUPT}} \times C + \beta_2 \times V.$$

To estimate and visualized the logistic regression models, we used the "glm" function and the "effects" (*53*) and "sjPlot" (*54*) packages for R.

**Data Analysis**

Robustness analysis
Some of our variables are correlated (e.g., methodological independence and prior knowledge independence; see Fig. S7), which is to be expected as they capture related dimensions of scientific knowledge production. We performed a multicollinearity test using the variance inflation factor (VIF). The variance inflation factors vary from low for variability in LINCS L1000 (VIF = 1.015), journal prominence (VIF = 1.135), and support in literature (VIF = 1.534) to moderate for centralization (VIF = 3.162), methodological independence (VIF = 3.752), prior knowledge independence (VIF = 4.681), and social independence (VIF = 4.900). We observe no predictor with high variance inflation factor, i.e., VIF ≥ 10. We removed the two variables with the highest VIF > 4 and refit our logistic regression model. In the refitted model, both support in the literature and decentralization of scientific communities remain strong and significant predictors of replication success (Fig. S10). Further, we verified that the effects of support from the literature and community centralization on claim replication success are not dominated by outliers. Recall the long-tailed distribution of findings per claim, with few claims receiving support from many published findings (Fig. 1A). We removed claims supported by 10 or more findings, amounting to 26 claims supported by 426 findings and found that support from the literature (OR 34.785; 95% CI: 12.518, 96.662, $P = 1.00\text{e-}11$) and community centralization (OR 0.335; 95% CI: 0.249, 0.451, $P = 5.57\text{e-}13$) remain strong and significant predictors of replication success, after controlling for biological and experimental variability in LINCS L1000 (OR 0.726; 95% CI: 0.429, 1.232; $P = 0.236$). Similarly, the distribution of findings per paper and per pair of papers is heterogeneous (Fig. S11). We removed the largest set of 796 drug-gene claims reported by a pair of papers and still found that the effect of support from the literature (OR 19.512; 95% CI: 7.193, 52.929, $P = 5.37\text{e-}09$) and centralization (OR 0.432; 95% CI: 0.287, 0.65, $P = 5.71\text{e-}05$) holds and is not explained by variability in LINCS L1000 (OR 0.574; 95% CI: 0.279, 1.181; $P = 0.131$).

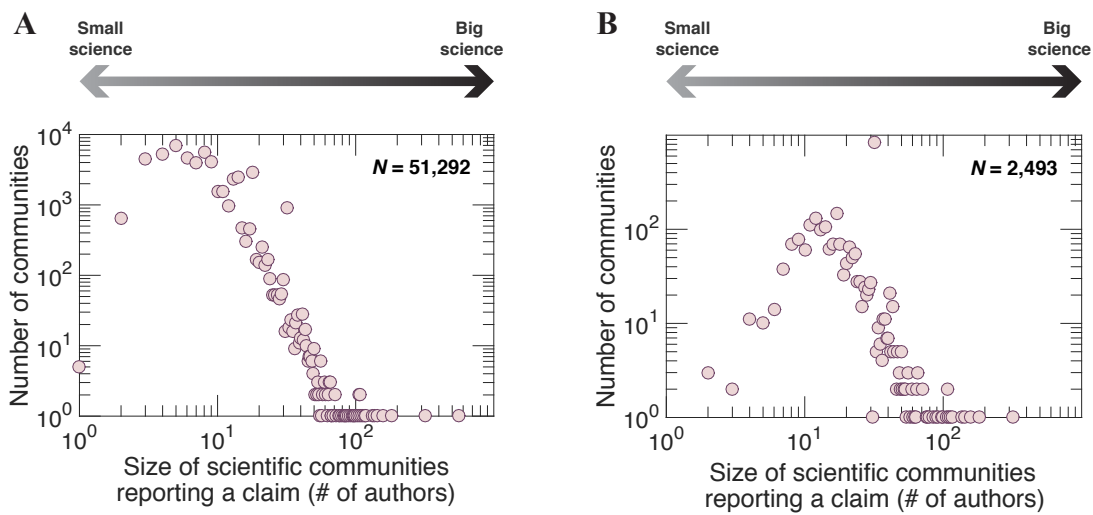

**Fig. S1. A continuum of "small" to "big" science communities defined as the total number of authors that have published papers reporting a drug-gene claim.** (A-B) Distribution of scientific communities of different sizes against the number of communities for the corpus of 51,292 and for the sub-corpus of 2,493 claims, respectively.

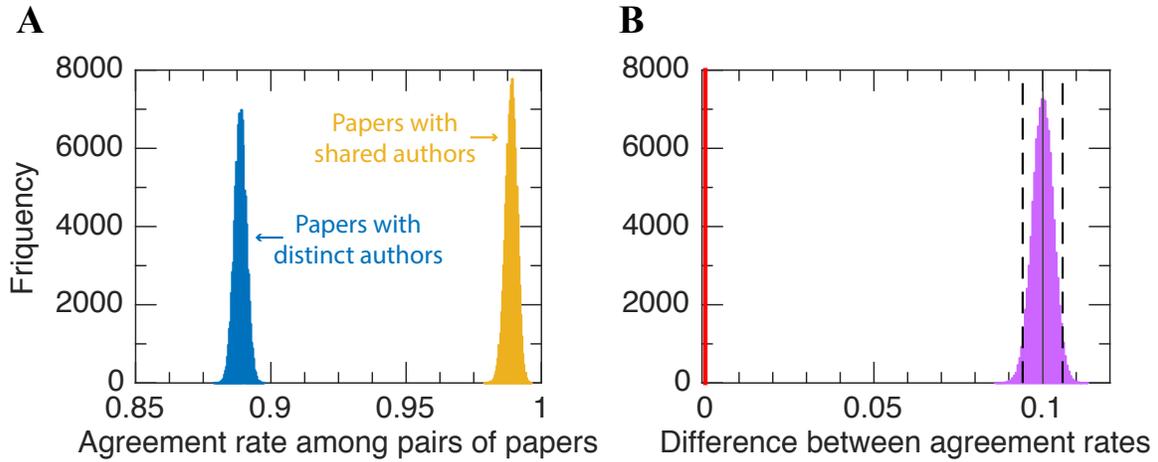

**Fig. S2. Publications that share authors are more likely to agree about the direction of a drug-gene interaction than publications with distinct authors, computed among pairs of papers reporting claims in the sub-corpus of 2,493 claims.** (**A**) Bootstrap samples of agreement (1 = agreement, 0 = disagreement) among pairs of papers that reported findings about the same drug-gene interaction depending on whether the pairs of papers shared (2,514 pairs of papers) or did not share author(s) in common (20,846 pairs of papers). (**B**) A two-tailed bootstrap test of difference between means indicates that papers with one or more authors in common have significantly higher agreement rates than pairs of papers with distinct authors (100,000 samples with replacement). The solid dark line indicates the sample mean, the dashed line indicates 95% CI, and the red line indicates the null hypothesis.

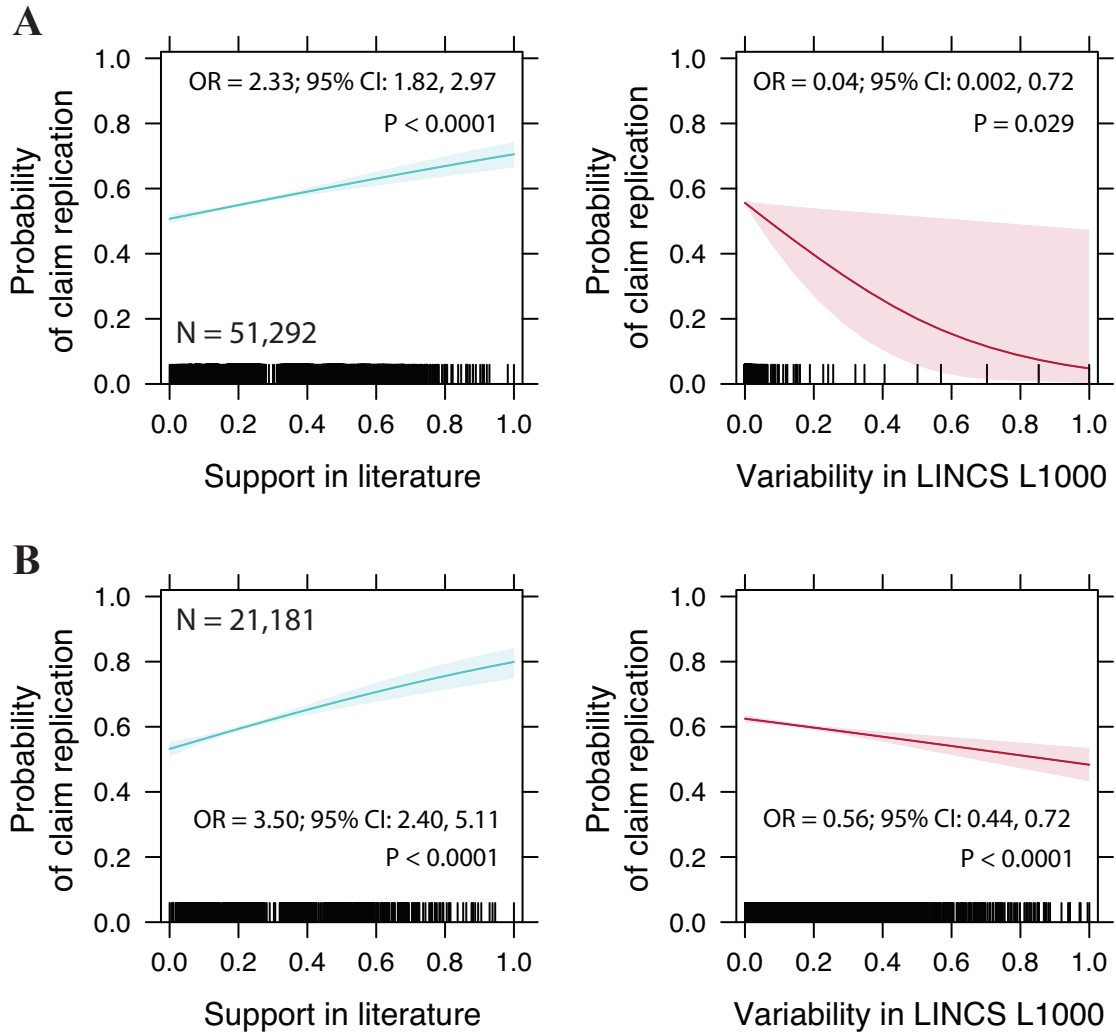

**Fig. S3. Replication success increases with claim's probability of support in the literature.** (**A**) Probability of claim replication estimated via a logistic regression model with replication as the response variable and support in the literature and experimental variability as predictors, for the whole corpus of 51,292 claims. (**B**) Same as **A** but for the subset of 21,181 significant drug-gene interactions in LINCS L1000. For each variable, we indicate odds ratios (OR), 95% confidence interval, and significance.

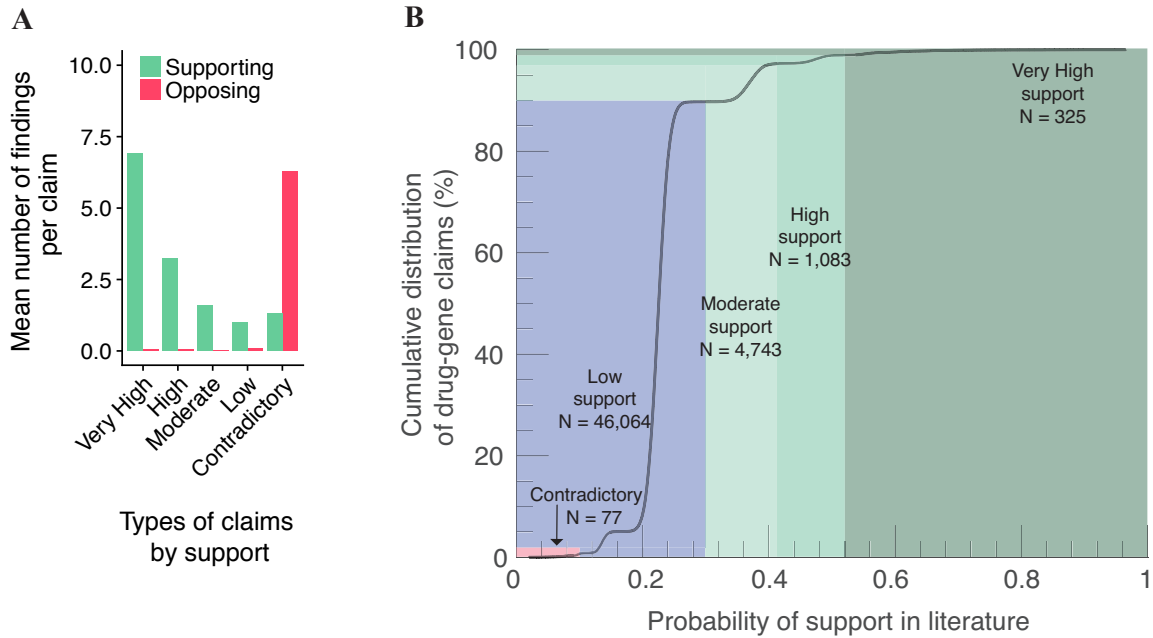

**Fig. S4. Description of claim types in the whole corpus of 51,292 claims.** (**A**) Mean number of findings supporting or opposing the direction of the effect for a scientific claim across our typology. (**B**) Cumulative distribution of the probability of support in the literature, including a schematic of our claim typology.

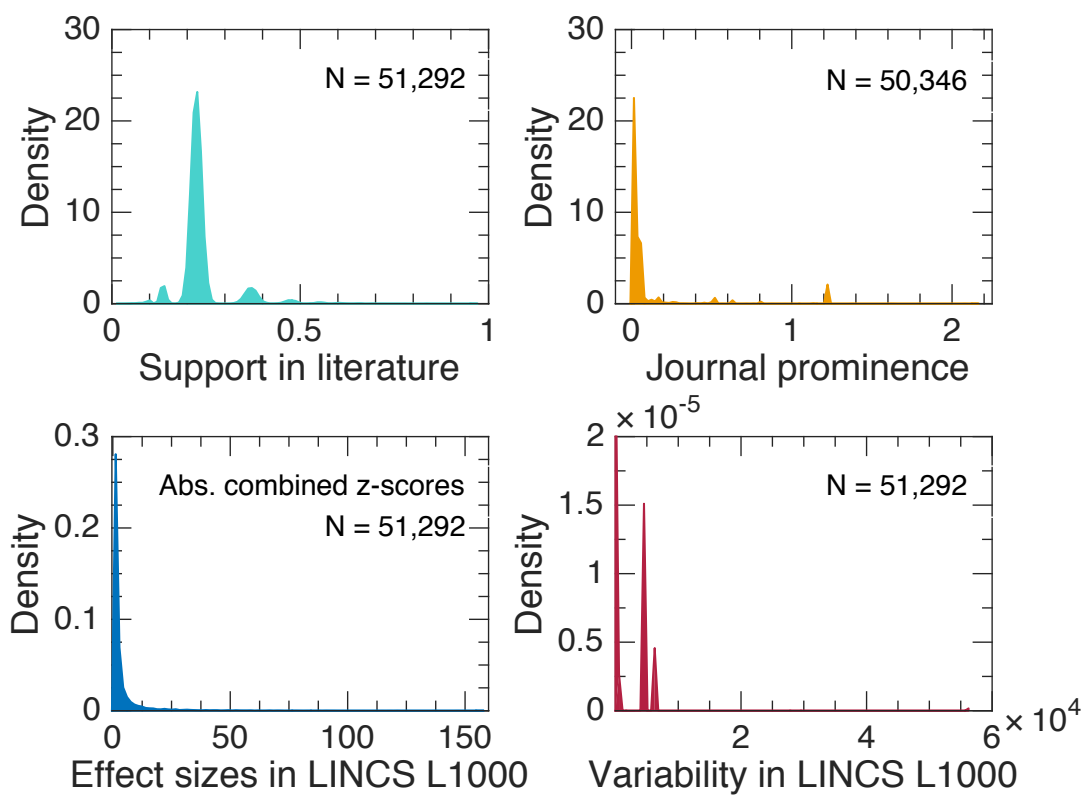

**Fig. S5. Estimates of probability density functions for variables of interest in our corpus using a normal kernel function.** We used the normal kernel function to estimate probability density functions.

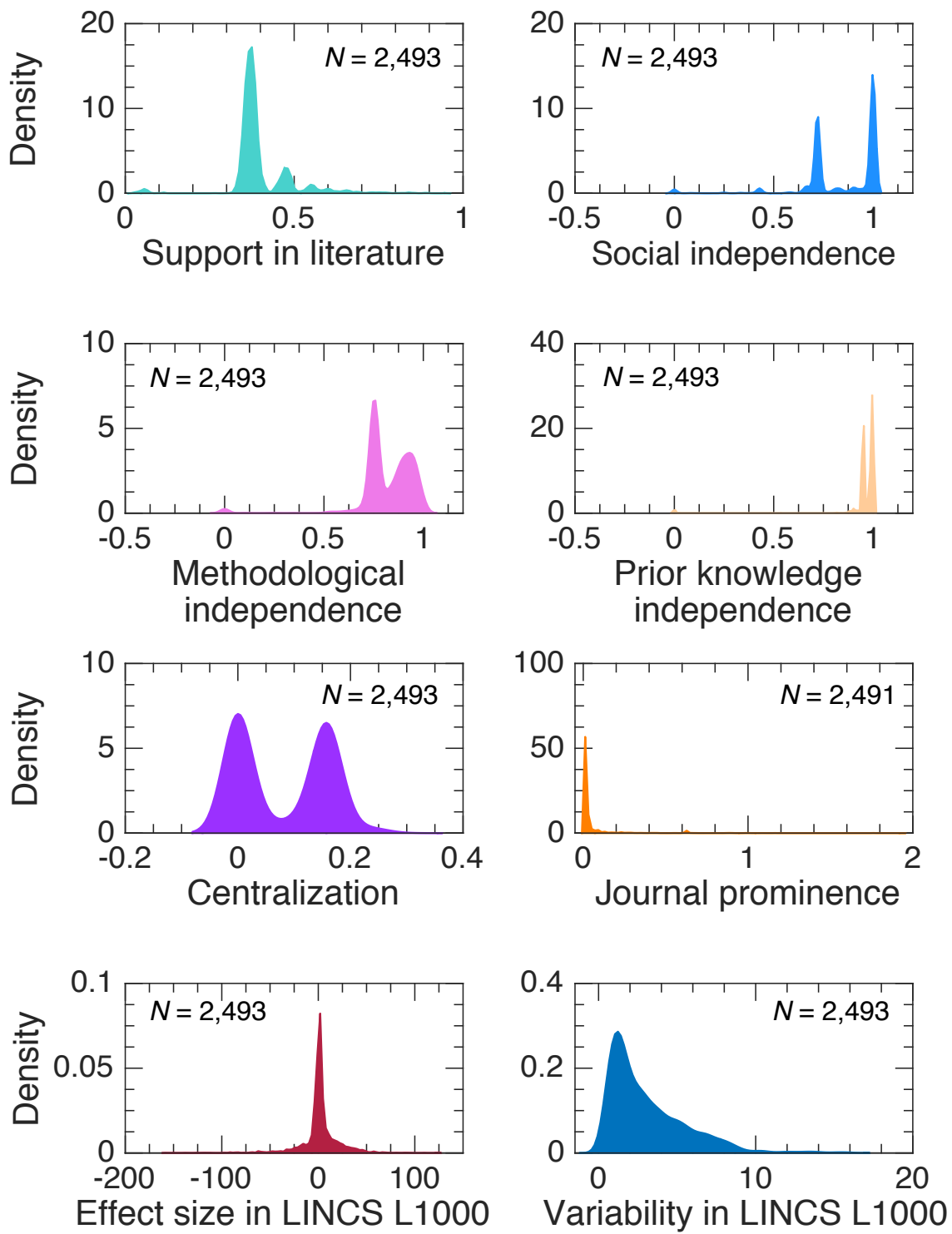

**Fig. S6. Estimates of probability density functions for our variables for the sub-corpus of claims with determined direction of the drug-gene effect in CTD and LINCS L1000.** As in Fig. S5, we used the normal kernel function to estimate probability density functions.

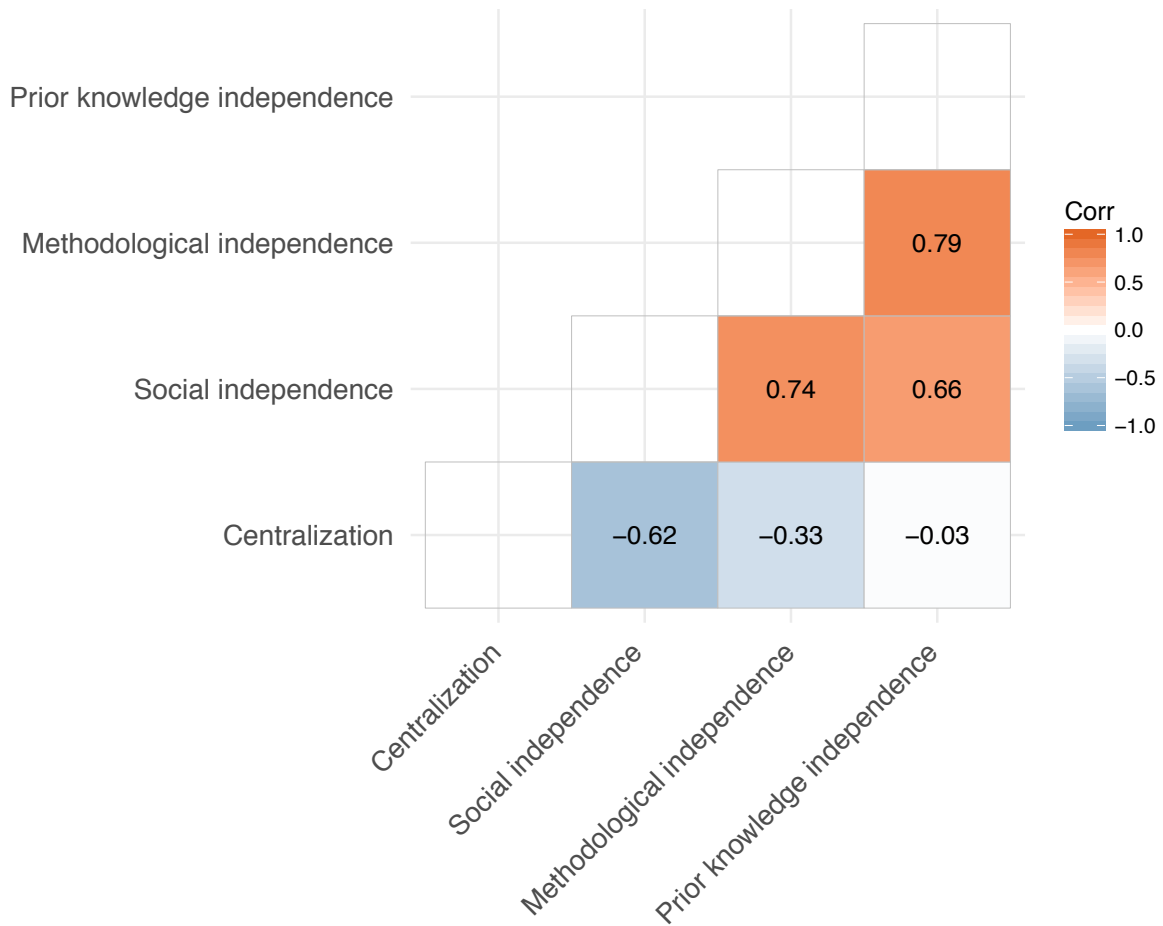

**Fig. S7. Pearson correlation coefficients between network indices.** Social, methodological, and prior knowledge (in)dependencies are positively correlated with each other and negatively correlated with the network centralization of scientific communities. $N = 2{,}493$ claims.

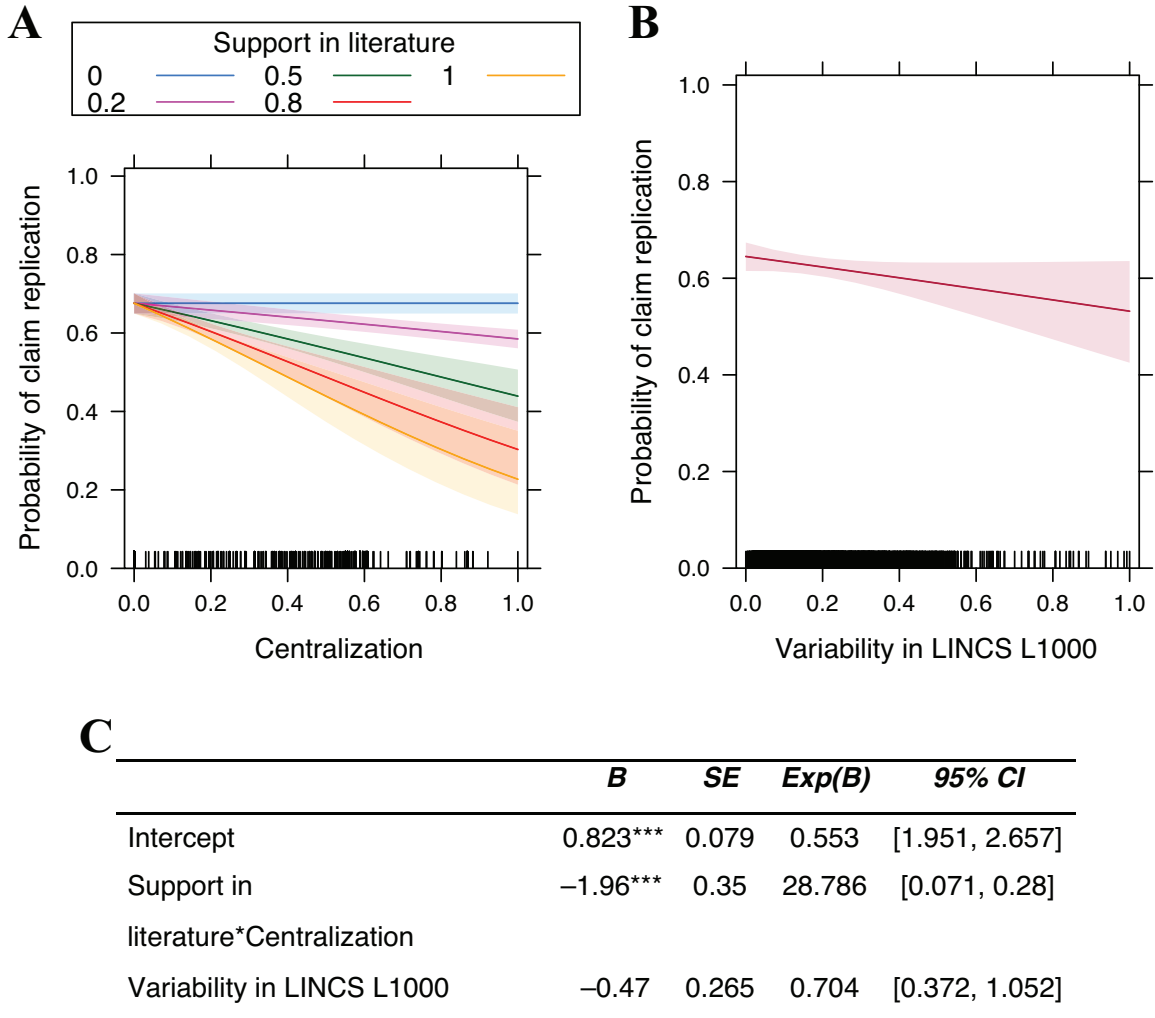

**Fig. S8. Claims reported by centralized communities less likely replicate. (A-C)**, Logistic interaction model with claim replication as the response variable regressed on support in the literature and centralization as interacting predictors, controlling for variability in LINCS L1000 across cell lines, durations, and dosages. Predictors are rescaled $\frac{x_i - \min(x)}{\max(x) - \min(x)}$ for comparability. $N = 2{,}493$ claims. *** $P < 0.001$; ** $P < 0.01$; * $P < 0.05$.

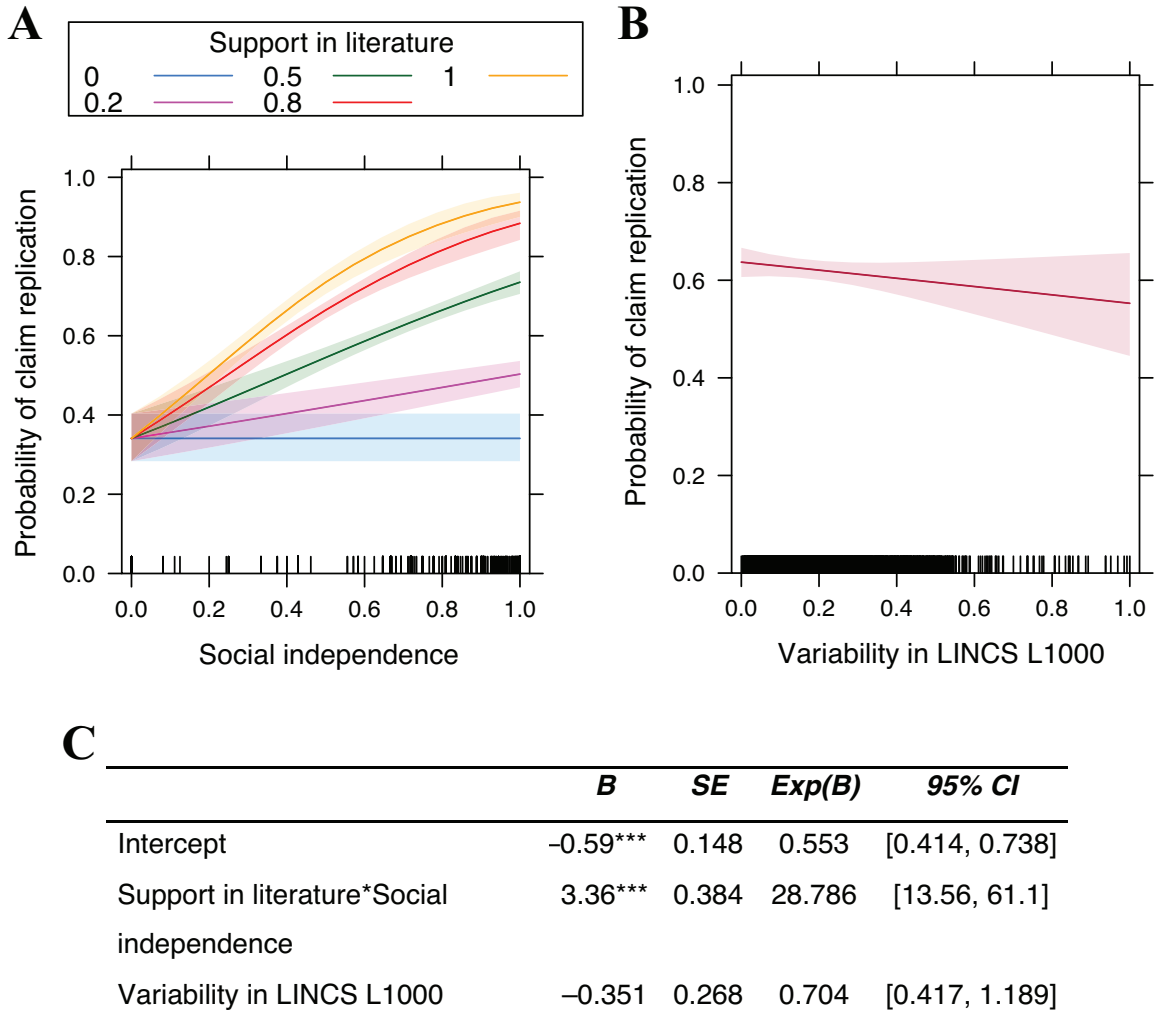

**Fig. S9. Claims reported by multiple socially independent teams more likely replicate.** (**A-C**) Logistic interaction model with claim replication as the response variable regressed on support from the literature and social independence as interacting predictors, controlling for variability in LINCS L1000 across cell lines, durations, and dosages. Predictors are rescaled $\frac{x_i - \min(x)}{\max(x) - \min(x)}$ for comparability. $N = 2{,}493$ claims. *** $P < 0.001$; ** $P < 0.01$; * $P < 0.05$.

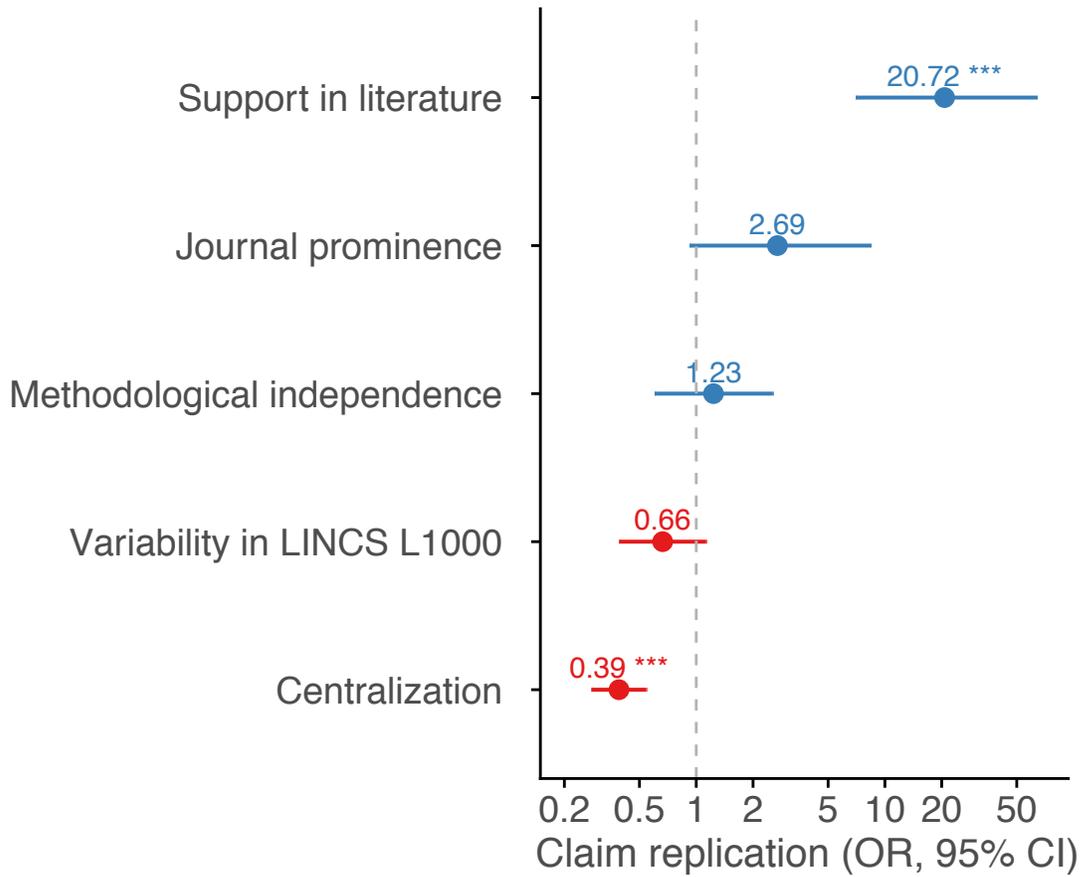

**Fig. S10. Support in the literature and decentralization of scientific communities remain strong and significant predictors of claim replication success after we account for multicollinearity.** Logistic model with replication as the response variable, including only predictors with relatively low variance inflation factor (VIF < 4). Predictors are rescaled $\frac{x_i - \min(x)}{\max(x) - \min(x)}$ for comparability. $N = 2{,}491$ claims. *** $P < 0.001$; ** $P < 0.01$; * $P < 0.05$.

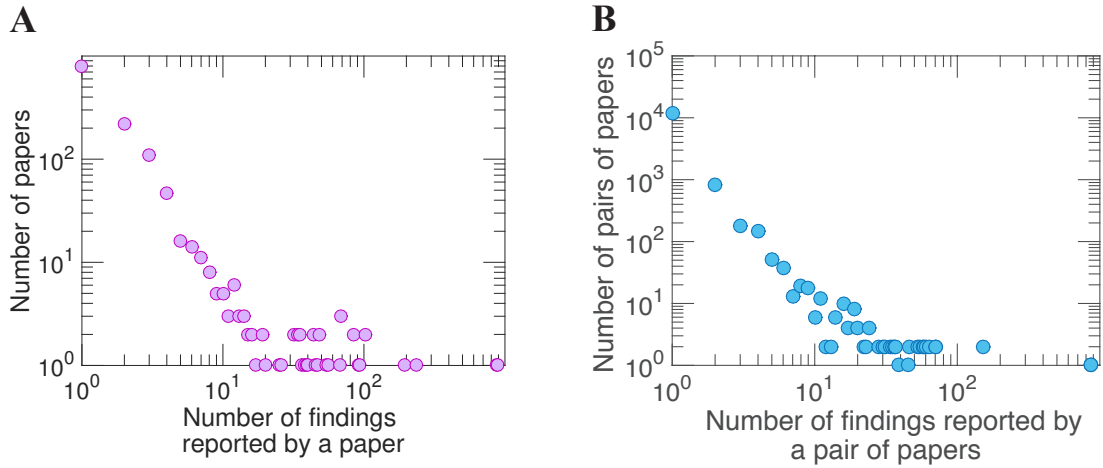

**Fig. S11. Papers and pairs of papers are differentiated by the number of findings they report (in the sub-corpus of 2,493 claims).** (**A**) Distribution of papers against the number of supporting findings they report. (**B**) Distribution of pairs of papers against the number of supporting findings they report.

| Variable | B | SE | Exp(B) | 95% CI |
|---|---|---|---|---|
| Intercept | –0.762*** | 0.195 | 0.47 | [0.318, 0.684] |
| Support in literature | 3.144*** | 0.479 | 23.2 | [9.078, 59.287] |
| | | | | |
| Intercept | 0.414*** | 0.046 | 1.513 | [1.384, 1.654] |
| Journal prominence | 2.379*** | 0.564 | 10.795 | [3.571, 32.638] |
| | | | | |
| Intercept | –1.053*** | 0.193 | 0.349 | [0.239, 0.509] |
| Social independence | 1.843*** | 0.224 | 6.312 | [4.069, 9.793] |
| | | | | |
| Intercept | –1.004*** | 0.255 | 0.366 | [0.222, 0.604] |
| Methodological independence | 1.841*** | 0.308 | 6.301 | [3.444, 11.527] |
| | | | | |
| Intercept | –1.14** | 0.376 | 0.32 | [0.153, 0.668] |
| Prior knowledge independence | 1.71*** | 0.388 | 5.53 | [2.584, 11.835] |
| | | | | |
| Intercept | 0.585*** | 0.065 | 1.794 | [1.579, 2.039] |
| Variability in LINCS L1000 | -0.444 | 0.264 | 0.641 | [0.382, 1.077] |
| | | | | |
| Intercept | 0.8*** | 0.062 | 2.224 | [1.972, 2.51] |
| Centralization | –1.018*** | 0.149 | 0.361 | [0.27, 0.484] |

**Table S1. Logistic regression models with claim replication *R* [Replicated = 1, Non-replicated = 0] as response variable and predictors modelled independently.** Predictors are rescaled $\frac{x_i - \min(x)}{\max(x) - \min(x)}$ for comparability. $N$ = 2,493 claims ($N$ = 2,491 in the Journal eigenfactor model); *** $P < 0.001$; ** $P < 0.01$; * $P < 0.05$.

| Variable | B | SE | Exp(B) | 95% CI |
| --- | --- | --- | --- | --- |
| Intercept | −0.693* | 0.352 | 0.5 | [0.251, 0.996] |
| Support in literature | 2.729*** | 0.595 | 15.316 | [4.772, 49.153] |
| Journal prominence | 1.019 | 0.563 | 2.769 | [0.918, 8.355] |
| Social independence | 0.632 | 0.486 | 1.881 | [0.725, 4.877] |
| Methodological independence | −0.183 | 0.585 | 0.833 | [0.265, 2.621] |
| Prior knowledge independence | −0.029 | 0.769 | 0.971 | [0.215, 4.383] |
| Variability in LINCS L1000 | −0.41 | 0.271 | 0.664 | [0.391, 1.128] |
| Centralization | −0.734** | 0.268 | 0.48 | [0.284, 0.812] |

**Table S2. Logistic regression models with claim replication *R* [Replicated = 1, Non-replicated = 0] as response variable and predictors modelled simultaneously.** Predictors are rescaled $\frac{x_i - \min(x)}{\max(x) - \min(x)}$ for comparability. $N = 2{,}491$ claims; *** $P < 0.001$; ** $P < 0.01$; * $P < 0.05$.